\documentstyle[12pt]{article}

\begin{document}
\thispagestyle{empty}

\def\ve#1{\mid #1\rangle}
\def\vc#1{\langle #1\mid}

\newcommand{\p}[1]{(\ref{#1})}
\newcommand{\be}{\begin{equation}}
\newcommand{\ee}{\end{equation}}
\newcommand{\sect}[1]{\setcounter{equation}{0}\section{#1}}

\renewcommand{\theequation}{\thesection.\arabic{equation}}

\newcommand{\vs}[1]{\rule[- #1 mm]{0mm}{#1 mm}}
\newcommand{\hs}[1]{\hspace{#1mm}}
\newcommand{\mb}[1]{\hs{5}\mbox{#1}\hs{5}}
\newcommand{\Db}{{\overline D}}
\newcommand{\bea}{\begin{eqnarray}}
\newcommand{\eea}{\end{eqnarray}}
\newcommand{\wt}[1]{\widetilde{#1}}
\newcommand{\und}[1]{\underline{#1}}
\newcommand{\ov}[1]{\overline{#1}}
\newcommand{\sm}[2]{\frac{\mbox{\footnotesize #1}\vs{-2}}
	   {\vs{-2}\mbox{\footnotesize #2}}}
\newcommand{\prt}{\partial}
\newcommand{\eps}{\epsilon}

\newcommand{\R}{\mbox{\rule{0.2mm}{2.8mm}\hspace{-1.5mm} R}}
\newcommand{\Z}{Z\hspace{-2mm}Z}

\newcommand{\cd}{{\cal D}}
\newcommand{\cg}{{\cal G}}
\newcommand{\ck}{{\cal K}}
\newcommand{\cw}{{\cal W}}

\newcommand{\vj}{\vec{J}}
\newcommand{\vl}{\vec{\lambda}}
\newcommand{\vz}{\vec{\sigma}}
\newcommand{\vt}{\vec{\tau}}
\newcommand{\vw}{\vec{W}}
\newcommand{\poiss}{\stackrel{\otimes}{,}}

\def\l#1#2{\raisebox{.2ex}{$\displaystyle
  \mathop{#1}^{{\scriptstyle #2}\rightarrow}$}}
\def\r#1#2{\raisebox{.2ex}{$\displaystyle
 \mathop{#1}^{\leftarrow {\scriptstyle #2}}$}}

\renewcommand{\thefootnote}{\fnsymbol{footnote}}
\newpage

\setcounter{page}{0}

\pagestyle{empty}

\vs{8}

\begin{center}
{\LARGE {\bf Graded Lie Algebras, Representation Theory,}}\\[0.6cm]
{\LARGE {\bf Integrable Mappings and Systems:}}\\[0.6cm]
{\LARGE {\bf Nonabelian Case}}\\[1cm]

\vs{8}

{\large A.N. Leznov$^{a,b,c}$}
{}~\\
\quad \\
{\em {~$~^{(a)}$ IIMAS-UNAM, Apartado Postal 20-726, Meïxico DF
01000, Meï
xico}}\\
{\em {~$~^{(b)}$ Institute for High Energy Physics,}}\\
{\em 142284 Protvino, Moscow Region, Russia}\\
{\em {~$~^{(c)}$ Bogoliubov Laboratory of Theoretical Physics,
JINR,}}\\
{\em 141980 Dubna, Moscow Region, Russia}

\end{center}

\vs{8}

\begin{abstract}

The exactly integrable systems connected with semisimple series $A$
for arbitrary grading are presented in explicit form. Their general
solutions
are expressed in terms of the matrix elements of various 
fundamental representations of $A_n$ groups. The simplest example of
such 
systems is the generalized Toda chain with the matrices of arbitrary 
dimensions in each point of the lattice.

\end{abstract}

\vfill

{\em E-Mail:\
leznov@ce.ifisicam.unam.mx}

\pagestyle{plain}

\renewcommand{\thefootnote}{\arabic{footnote}}

\setcounter{footnote}{0}

\section{Introduction}

More than fifteen years ago it became clear that with each graded Lie
algebra
it is possible to connect the exactly integrable system \cite{l3}.
The 
general solution of these systems may be obtained in a finite number 
of algebraic operations (including the quadratures). But at that time 
the connection of such kind systems with evolution type equations, 
integrated by inverse scattering method \cite{L4}, was not observed
in 
explicit form.

At the present moment situation is as follows: definite class of 
exactly integrable systems is equivalent to integrable mappings, with
respect 
to which evolution type equations are invariant \cite{L1}. This
invariance 
allows to obtain the explicit particular solutions of evolution type
equations 
(including the soliton type subclass). Thus the description of the
most large 
class of integrable mappings simultaneously solve the problem of the
constructing and integrating of evolution type systems and equations. 

The theory of representation of semisimple algebras at the present
time
in many aspects is the closed one and so it is natural to try to
represent in explicit form all exactly integrable systems (integrable 
mappings) connected with arbitrary semisimple algebras together with
the
arbitrary grading in it at least in the framework of the construction 
proposed in \cite{l3}. 

In a series of the recent author's papers \cite{LM1}, \cite{LM2},
\cite{LM3} 
this program was partially realized. The connection between the
graded
algebras, integrable mappings, exactly integrable systems and
evolution
type equations on the example of $A_n$ semisimple series and main
grading in 
it ( see in this connection \cite{UT}) was demonstrated in
\cite{LM1}. For 
arbitrary semisimple algebras ( only for the case of the main 
grading ) the explicit form of exactly integrable systems together
with their 
general solutions have been found in \cite{LM2}. And at last the same
problem  
for the case of the semisimple algebras of the second rank but for
arbitrary
choice of the grading in them was solved in \cite{LM3}.

The main goal of the present paper is to give solution of this
problem for
the case of $A_n$ series with arbitrary grading in it. This case
includes particularly ( as the simplest one) the generalization of
matrix
Toda chain with unknown matrix functions of the different dimension
in each 
point of the lattice.  

Few words about the notations. In construction below the following
main 
ingredients will be used: semisimple algebra, grading in it, maximal
numbers 
of graded subspaces (positive and negative) exploited and
nonequivalent 
canonical forms to which may be performed with the help of the gauge 
transformation, the subspaces of maximal graded indexes. 
In connection with this we call the exactly integrable system under 
consideration by the letter of corresponding semisimple series $G$,
as a 
function on the grading vector (consisting of the zeros and unites
in 
arbitrary order), by two natural numbers $m_1,m_2$- the maximal
graded indexes 
of the subspaces involved and indexes $C, \bar C$- the
characteristics of 
above mentioned canonical forms ( the fine structure), the choice of
which may 
lead to the systems with the absolutely different symmetry type
\cite{DLS}. 
In these notations usual Toda chain looks as the following
$G((1,...1);1,1;
(1,..1),(1,..1))$ and really is the particular simplest example,
compare with 
the rich possibilities of the general construction.  

The paper is organized as follows. In section 2 we present (as a rule
without
any proofs) necessary facts and formulae from the representation
theory of the
semisimple algebras and groups. Section three is devoted to
description
of the general construction and mathematical tricks and methods used
in
the main sections. In the 4 section the concrete example of
generalised 
matrix Toda chain with the matrices of arbitrary dimension in each
point of 
the lattice is considered.  In section 5 we get rid of the
restriction of 
using only the graded subspaces with $0$ and $\pm 1$ graded indexes
and 
represent corresponding exactly integrable system. Concluding remarks
and 
perspectives for further investigation are concentrated in section 6. 

\section{Necessary facts from the representation theory of the
semisimple
algebras and groups}

Let ${\cal} G$ be some arbitrary finite dimensional graded Lie
algebra 
\footnote{We make no difference between algebra and super-algebra
cases, 
recalling only that even (odd) elements of the super-algebras are
always 
multiplied by even (odd) elements of the Grassmann space.}.
This means that $\cal G$ may be represented as a direct sum of
subspaces with
the different grading indexes
\begin{equation}
  {\cal G}=\left(\oplus^{N_-}_{k=1} {\cal G}_{-\frac{k}{2}}\right)
  {\cal G}_0 \left(\oplus^{N_+}_{k=1}{\cal G}_{\frac{k}{2}}\right).
\label{GR}
\end{equation}

The generators with the integer grading indexes are called bosonic,
while
with half-integer the fermionic ones. Positive (negative) grading
corresponds to upper (lower) triangular matrices.

The grading operator $H$ for arbitrary semisimple algebra may be
represented
as linear combination of elements of commutative Cartan subalgebra,
taking
the unity or zero values on the generators of the simple roots 
in the form:
\begin{equation}
H={\sum}^r_{i=1} (K^{-1}c)_i h_{i}
\label{cartan1}
\end{equation}
Here $K^{-1}$ is the inverse Cartan matrix $K^{-1}K=KK^{-1}=I$
and $c$ is the column consisting from zeros and unities in arbitrary
order.
Under the main grading all $c_i=1$ and in this case
$(K^{-1}c)_i=\sum_{j=1}
^r K^{-1}_{i,j}$, $r$ is the rank of the algebra.

As usually the generators of the simple roots $X^{\pm}_i$ (raising,
or
lowering
operators) and Cartan elements $h_i$  satisfy the system of
commutation
relations:
\begin{eqnarray}
[h_i , h_j]=0, \quad [h_i,X^{\pm}_j]=\pm K_{j,i}X^{\pm}_j, \quad
[X^{+}_i,X^{-}_j\}={\delta}_{i,j} h_j, \quad (1 \leq i,j \leq r),
\label{aa6}
\end{eqnarray}
where $K_{ij}$ are the elements of Cartan matrix, and the brackets
$[,\}$ 
denote the graded commutator, $r$- is the rank of the algebra.

The highest vector $\ve{j}$ ($\vc{j} \equiv \ve{j}^{\dagger}$) of
the $j$--th fundamental representation possesses the following
properties:
\begin{eqnarray}
X^{+}_i\ve{j}=0, \quad h_i\ve{j}={\delta}_{i,j}\ve{j}, \quad
\vc{j}\ve{j}=1.
\label{high}
\end{eqnarray}
The representation is exhibited by repeated applications of the
lowering operators $X^{-}_i$ to the $\ve{j}$ and
extracting all linear-independent vectors with non-zero norm. Its
first
few basis vectors are
\begin{eqnarray}
&& \ve{j}, \quad X^{-}_j\ve{j},  \quad X^{-}_i X^{-}_j\ve{j},\quad
i\neq j,
\quad K_{i,j}\neq 0
\label{vectors}
\end{eqnarray}

In the fundamental representations, matrix elements of the
group element $G$ satisfy the following important identity
\footnote{Let us remind the definition of the
superdeterminant, $sdet \left(\begin{array}{cc} A, & B \\ C, & D
\end{array}\right) \equiv det (A-BD^{-1}C ) (det D)^{-1}$.}
\cite{MT1}
\begin{eqnarray}
sdet \left(\begin{array}{cc} \vc{j}X_j^+GX_j^-\ve{j}, &
\vc{j}X_j^+G\ve{j} \\ \vc{j}GX_j^-\ve{j}, & \vc{j} G \ve{j}
\end{array}\right) = {\prod}^r_{i=1}\vc{i} G \ve{i}^{-K_{ji}},
\label{recrel}
\end{eqnarray}
where $K_{ji}$ are the elements of the Cartan matrix. The identity
(\ref
{recrel}) represents the generalization of the famous Jacobi identity
connecting determinants of (n-1), n and (n+1) orders of some special
matrices to the case of arbitrary semisimple Lee super-group. As we
will see
in the next sections, this identity is so important at the deriving
of the
exactly integrable systems, that one can even say that it is
responsible for
their existence. We conserve for (\ref{recrel}) the name of the first
Jacobi
identity. Besides (\ref{recrel}), there exists no less important
independent identity \cite{UT}:
$$
(-1)^P K_{i,j} {\vc{j}X_j^+X_i^+ G \ve{j}\over \vc{j} G \ve{j}}+
K_{j,i} {\vc{i}X_i^+X_j^+ G \ve{i}\over \vc{i} G \ve{i}}+
$$
\begin{equation}
K_{ij}K_{j,i}(-1)^{jP}{\vc{j}X_j^+ G \ve{j}\over \vc{j} G
\ve{j}}{\vc{i}X_i^+
G \ve{i}\over \vc{i} G \ve{i}}=0 ,\quad i\neq j \label{J2}
\end{equation}
which will be called as a second Jacobi identity. This identity is
responsible (in the above sense) for the fact of existence of
hierarchy
of integrable systems, each one of which is invariant with respect to
transformations of the integrable mapping, connected with each
exactly
integrable system.

Either from (\ref{recrel}) or from (\ref{J2}) it is possible to
construct
many useful recurrent relations which will be used in the further
consideration.

Keeping in the mind the importance of the Jacobi identities
(\ref{recrel}) 
and (\ref{J2}) for further consideration for convenience of the
reader we
present below briefly a proof of (\ref{recrel}).

Let us consider the left hand side of (\ref{recrel}) as a function on
the
group, where $G$ is its arbitrary element. The action on $G$ in the
deffinite
representation $l$ of the operators of the right (left) regular
representation is by definition as follows:
\begin{equation}
M_{left}(\tilde M_{right}) G= M_l G (\tilde M_l)\label{AM}
\end{equation}
where now $M_l,\tilde M_l$ are generators (the matrices of
corresponding 
dimension) of the shifts on the group in a given $l$ representation.
Now let 
us act on left hand side of (\ref{recrel}) by arbitrary generator
of simple 
positive root $(X^+_s)_r$.
This action is equivalent to differentiation and so it is necessary
to
act consequently on the first and the second columns of the matrix
(\ref{recrel}) with summation of the results. The action on the
second one 
is always equal to zero as a corollary of the definition of the
highest state 
vector (\ref{high}).
The action on the first column is different from zero only in the
case $s=j$.
But in this case using the same definition of the highest state
vector
we conclude that in the result of differentiation of the first column
it
becomes equal to the second one with the zero final result. So as the
function on the group the left hand side of (\ref{recrel}) is also
proportional to the highest vector of some other representation ( or
linear
combination of such kind functions). The highest vector of the
irreducible 
representation is uniquely defined by the values, which Cartan
generators take 
on it. If Cartan generators take on the highest 
vector values $V(h_i)=l_i$, then the last one may be uniquely
represented in 
the form:
\begin{equation}
\vc{l} G \ve{l}=C \prod_{i=1}^r(\vc{i} G \ve{i})^{l_i}\label{HVD}
\end{equation}
Calculating the values of Cartan generators on the left hand side of 
equation (\ref{recrel}) (both left and right with the same result)
and using 
the last comment about the form of the highest vector we prove
(\ref{recrel}) 
( the constant C=1, as one can see by putting $G=1$ and comparing the
terms
on the both sides).

The second Jacobi identity can be proved with the help of the
arguments of the
same kind \cite{UT}.

The following generalization of the first Jacobi identity will be
very
important in calculations, connected with nonabelian gradings.

Let $\ve{\alpha}$ be the basis vectors of some representation 
rigorously in the order of increasing of the number of the lowering 
generators (see (\ref{vectors})). We assume also that under the
action of 
generators of arbitrary positive simple root on each basis vector 
there arises the linear combination of the previous ones.

Then the principal minors of arbitrary order of the matrix ($G$ is
arbitrary 
element of semisimple group):
$$
G^{\alpha}=\vc{\alpha} G  \ve{\alphaï} 
$$
are annihilated by all generators of positive root from right and
negative 
ones from left.

Indeed this action is equivalent to differentiation and so it is 
necessary to act separately on each column (line) of the minor's
matrix with 
further summed the results. But action of generator of the positive 
simple root on state vector with given number of the lowering
operators 
perform it to the state vector with number of lowering operators on
unity 
less. Thus in all cases the determinant arises with the linear
dependence 
between it's lines or columns always with the zero result.
The generators of Cartan subalgebra obviously take the definite
values on 
the the minors of such kind and if the corresponding values of them
are 
$l^s_i$, then it is possible to write in correspondence with
(\ref{HVD})
the equality:
\begin{equation}
Min_s ( G^{\alpha})= C_{\alpha,s} \prod_{i=1}^r \vc{i} G
\ve{i}^{l^s_i}
\label{GYI}
\end{equation}
and corresponding constant may be determined as it was described
above.

\section{General construction and technique of computation}

The grading of a semisimple algebra is defined by the values, which
grading
operator $H$ takes on the simple roots of the algebra. This values
may be only
zero and unity ones in arbitrary order. These facts are encoded in
the 
equations:
$$
[H, X^{\pm}_i]=\pm X^{\pm}_i,\quad H=\sum_1^r (K^{-1}c)_i h_i,\quad
c_i=1,0
$$

On the level of Dynkin's diagrams the grading can be introduced by
using two 
colors for its dot's: black for simple roots with $c_i=1$ and red
ones 
for the roots with $c_i=0$.\footnote{Red and black colors are in deep 
connection with two levels of Johnny Walker theory.}
With each consequent sequence of the red (simple) roots
it is possible to connect the corresponding semisimple algebra (
subalgebra of the
the initial one). All such algebras are obviously mutually
commutative
and 
belong to zero graded subspace. To zero graded subspace belong also
Cartan 
elements of the all black roots. We conserve the usual numeration of
the dots 
of Dynkinïs diagrams and all red algebras will be distinguished by
the index 
of its first root $m_s$. The rank of $m_s$-th semisimple algebra will
be 
denoted as $R_s$. So
$X^{\pm}_{m_s},X^{\pm}_{m_s+1},....X^{\pm}_{m_s+R_s-1}$ 
are the system of the simple roots of $m_s$ red algebra.

After these preliminary comments we pass to description of the
general 
construction \cite{l3}.
Let two group valued functions $ M^+(y), M^-(x)$ are solutions of
$S$- matrix type
equations:
\begin{equation}
M^+_y=((\sum_0^{m_2} B^{(+s}(y)) M^+\equiv (B^{(0}+L^+)M^+\quad
M^-_x=M^-(\sum_0^{m_1} A^{(-s}(x))\equiv M^-(A^{(0}+L^-) \label{I}
\end{equation}
where $B^{(+s}(y), A^{(-s}(x)$ take values correspondingly in $\pm s$
graded subspaces.
 $s=0, 1, 2,...m_{1,2}$. In each finite dimensional representation
 $B^{(+s}
(y), A^{(-s}(x)$ are upper (lower) triangular matrices and so
equations 
(\ref{I}) are integrated in quadratures.

The key role in what follows plays the composite group valued
function $K$:
\begin{equation}
K=M^+ M^-. \label{II}
\end{equation}
It turns out that matrix elements of $K$ in various fundamental 
representations are connected by closed systems of equivalent
relations, 
which can be interpreted as exactly integrable system with known
general 
solution.

Now we pass to the describing of the necessary calculation methods to
prove 
this proposition.

First of all let us calculate the second mixed derivative $(\ln
\vc{i} K 
\ve{i})_{x,y}$, where index $i$ belongs to black dot of Dynkins
diagram. 
We have consequently:
\begin{equation}
(\ln \vc{i} K \ve{i})_x= {\vc{i} K (A^0+L^-) \ve{i}\over \vc{i} K
\ve{i}}=
A^0_i(x)+{\vc{i} K L^- \ve{i}\over \vc{i} K \ve{i}}
\label{MI}
\end{equation}
Indeed, $ K_x=M^+(y)M^-_x(x)=K (A^0+L^-)$ as a corollary of equation
for $M^-$. All red 
components of $A^0$ under the action on the black highest vector
state 
$\ve{i}$ lead to a zero result in connection with (\ref{high}). The
action of 
Cartan elements of the black roots satisfy the condition $h_j
\ve{i}=\delta_
{i,j} \ve{i}$ and so only coefficient of $h_i$ remains in the final
result 
(\ref{MI}).

Further differentiation (\ref{MI}) with respect to $y$, with the 
help of the arguments above, leads to following result:
\begin{equation}
(\ln \vc{i} K \ve{i})_{x,y}=\vc{i} K \ve{i}^{-2}
\pmatrix{ \vc{i} K \ve{i} &  \vc{i} K L^- \ve{i} \cr
	 \vc{i} L^+ K \ve{i} &  \vc{i} L^+ K L^- \ve{i} \cr}.\label{NA}
\end{equation}                                       

Applying (\ref{AM})  of the previous section to the left hand side of 
(\ref{NA}), we obtain finally:
\begin{equation}
(\ln \vc{i} K \ve{i})_{x,y}=L^-_r L^+_l \ln \vc{i} K \ve{i}
\label{ARR}
\end{equation}

And thus the problem of the calculation of the mixed second
derivative is 
reduced to the pure algebraic manipulations on the level of 
representation theory of semisimple algebras and groups. Further
evaluation of 
(\ref{ARR}) is connected with repeated applications of the first 
(\ref{recrel}) and second (\ref{J2}) Jacobi identities as it will be
clear 
from the material of the next section.

As was mentioned above the red algebras of the zero order graded
subspace in 
general case are not commutative ones and this leads to additional 
computational difficulties. Let us denote by $ \ve{m_i}$
the highest vector of $m_i$-th fundamental representation of the
initial 
algebra. Of course, $\ve{m_i}$ is simultaneously the highest vector
of the 
first fundamental representation of the $m_i$ red algebra. Let
$\vc{\alpha_i}, 
\ve{\alphaï_i}$ be the basis vectors of the first fundamental
representation 
(this restriction is not essential) of $m_i$-th red algebra and let
us
consider the matrix elements of element $K$ in this basis.
$R_i+1\times R_i+1$ matrix ( $R_i+1$ is the dimension of the first
fundamental 
representation), with the matrix elements  $\vc{\alpha_i} K
\ve{\alpha^ï_i}$ 
will denoted by a single symbol $u_i$ ( index $i$ takes values from
one to the 
number of the red algebras, which is the function of the chosen
grading).

For derivatives of matrix elements of such constructed matrix we have 
consequently (index $i$ we omit for a moment):
\begin{equation}
\vc{\alpha} u_x \ve{\alphaï}=\vc{\alpha} K (A^0+L^-)
\ve{\alphaï}=\sum_
{\gamma} \vc{\alpha} K  \ve{\gamma}\vc{\gamma} I A^0 \ve{\alphaï}
+\vc{\alpha} K L^- \ve{\beta}\label{MCD}
\end{equation}
Or in the equivalent form:
$$
u^{-1} u_x= A^0(x)+u^{-1} \vc{} K L^- \ve{}
$$
Further differentiation with respect to variable $y$ leads to:
$$
\vc{}((u^{-1} u_x)_y\ve{}=u^{-1} \vc{} (B^0+L^+) K L^-
\ve{}-u^{-1}\vc{} 
(B^0+L^+) K \ve{} u^{-1}\vc{} K L^-\ve{}=
$$
\begin{equation}
{}\label{MC}
\end{equation}
$$
u^{-1}(\vc{} L^+ K L^- \ve{}-\vc{} L^+ K \ve{} u^{-1} \vc{} K
L^-\ve{})
$$
The last expression with the help of standard transformations may be
brought 
to the form of the ratio of the two determinants correspondingly of
the 
$R_i+2$ and $R_i+1$ orders:
\begin{equation}
\vc{} u (u^{-1} u_x)_y \ve{}={Det_{R_i+2}\pmatrix{ u & K L^-\ve{} \cr
\vc{} L^+ K  & \vc{} L^+ K  L^- \ve{} \cr}\over Det_{R_i+1}(u)}
\label{MCC}
\end{equation}

The key role in the process of the rediscovering of the last
expression 
plays the generalised Jacobi identity (\ref{GYI}).

\section{Generalized matrix Toda chain}

The title of this section is decoded in the notations of introduction
as 
follows: 
$$
A_n((0,..0,1),(0...0,1),(0......0,1),(0...0));1,1; C,\bar C). 
$$
The zero subspace consists from the mutual commutative $m_i$ red
algebras 
$A_{R_i}$ and Cartan elements $h_{m_i+R_i}$, which are noncommutative
with 
the first and the last simples roots following after and previously 
red algebras. The meaning of $C,\bar C$ will be clarified in what 
follows. In the case, when $m_i=M$ we come back to the case of so
called 
matrix Toda chain, considered from the different point of view in the
papers 
\cite{MT1},\cite{MT2},\cite{MT3}.

The system of $(R_i+1)$ basis vectors of the first fundamental
representation 
of the $m_i$ red algebra $A_{R_i}$ is the following (index $i$ will
be 
omitted on a time):
\begin{equation}
\ve{m},
X^-_m\ve{m},X^-_{m+1}X^-_m\ve{m},.....X^-_{m+R-1}....X^-_m\ve{m}
\label{SV1}
\end{equation}
The system of $(R+1)$ basis vectors of its last fundamental
representation is:
\begin{equation}
\ve{m+R-1}, X^-_{m+R-1}\ve{m+R-1},....X^-_m....X^-_{m+R-1}\ve{m+R-1}
\label{SV2}
\end{equation}
It will be suitable sometimes to enumerate the basis (\ref{SV1}) with
the help 
of the natural positive numbers from $1$ up to $(R+1)$ from left up
to right 
and basis (\ref{SV2}) in the same manner but in the opposite
direction. 
Thus:
$$
\vc{1}\equiv \vc{m},\quad \vc{2}\equiv \vc{m} X^+_m,\quad
\vc{j}\equiv \vc{m} 
X^+_m...X^+_{m+j-2},\quad (X^+_mX^+_{m-1}=1).
$$

Of course, these sequences of basis vectors are only a part of the
same for 
the $m_i$ and $m_i+R_i-1$ fundamental representations of the initial 
$A$ algebra.

In what follows the notation $\bar m_i\equiv m_i+R_i-1$,
$m_{i+1}=\bar m_i+2$ 
will be used. And also:
$$
b_i\equiv \bar m_i+1\quad <a>=\vc{a} K \ve{a}
$$

Two $ (R+1)\times (R+1)$- matrix valued functions $u_m$ and $u_{\bar
m}$ 
constructed from the matrix elements of arbitrary group element $G$
in two 
bases above will be important for the further consideration.

As a direct corollary of the generalised Jacobi identity (\ref{GYI})
the 
following proposition takes place: 

Two $(R_i+1)\times (R_i+1)$ matrix functions
$$
y_i={u_{m_i}\over < b_{i-1} >},\quad z_i={u_{\bar m_i}\over 
< b_i >}
$$
are connected by the relation:
\begin{equation}
y_i^{-1}=t_i z^T_i t_i^{-1}\label{MTE}
\end{equation}
where $t_i$ is $(R_i+1)\times (R_i+1)$ matrix with non zero matrix
elements
only on the main antidiagonal, consisting from $\pm 1$ in 
the exchange order, $T$ is the sign of transposition.

First of all let us calculate the determinants of $u_m,u_{\bar m}$.
The
matrices of both determinants satisfy conditions of generalised
Jacobi identity of the end of the second section. So it is necessary
only to 
calculate the summed values of Cartan generators  $Vh$ on the
sequence 
of the state vectors (\ref{SV1}) and (\ref{SV2}).
In the first case nonzero are only $Vh_{m-1}=R,Vh_{m+R}=1$ in the 
second one $Vh_{m-1}=1, Vh_{\bar m+1}=R$ and so we conclude:
\begin{equation}
Det (u_{m_i})=< b_{i-1} >^{R_i}< b_i > ,\quad Det (u_{\bar m_i})=<
b_i >^
{R_i}< b_{i-1} > 
\label{DU}
\end{equation}
and so really $Det y_i\times Det z_i=1$.

To prove the proposition (\ref{MTE}) in the whole volume it is
sufficient to 
present the matrix elements of inverse matrix as a ratio of
corresponding
minors of $R_i$ order matrix $u$ to it determinant and apply the
generalized 
Jacobi identity (\ref{GYI}).

For further calculation for us will be important more detailed
knowledge of 
the graded structure of the initial algebra, which we present with
the help 
the diagram below and further comments to it.
$$
\begin{tabular}{|r|l|c|}\hline
{}{}{}{}{} &{}{}&{}{}{}{}{}{}\\ 
$m_{i-1}$  &$I^{+1}_{i-1}$ & $I^{+2}_{i-1}$ \cr 
	   &      &       \\ \hline 
	   &      &        \\
$I^{-1}_{i-1}$ & $ m_i$ & $I^{+1}_i$ \cr
	   &      &      \\  \hline
	   &      &       \\  
$I^{-2}_{i-1}$ & $I^{-1}_i$ & $m_{i+1}$ \cr
	   &      &       \\  \hline
\end{tabular}
$$

Here quadratic blocks are the root spaces of $m_i$ red algebras;
$I^{\pm}_i$
are components of subspace with graded indexes $\pm 1$. They are
described 
with the help of rectangular $(R_i+1)\times (R_{i+1}+1)$ and
$(R_{i+1}+1)
\times (R_i+1)$ matrices, respectively. $I^{\pm}=\sum_i I^{\pm}_i$,
$I^-_i=
(I^+_i)^T$. All components of $\pm 1$ graded subspace, except of
$I^{\pm}_i,
I^{\pm}_{i-1}$ are mutually commutative with root space of $m_i$ red
algebra
and lead to zero result under the action on the basis vectors
(\ref{SV1}) and
(\ref{SV2}).

It will be suitable to denote each composite root of the initial $A$
algebra 
by the indexes of its first and last simple roots, from which it is
composed ( so called tensor notations): 
$$
(i,j)^+=[X^+_i[X^+_{i+1}[....[X^+_{j-1},X^+_i]..]], i\leq j, 
(i,i)^+=X^+_i, (i,{i-1})^+=1, (i,j)^-=((i,j)^+)^T
$$

The components of $L^{\pm}$, taking values in  $I^{\pm}_i$, have the
form: 
\begin{equation}
L^+_i=\sum_{s=1}^{R_i+1} \sum_{j=1}^{R_{i+1}+1}
\bar P^i_{s,j}(y) [(m_i+s-1,\bar m_i)^+[(\bar m_i+1,\bar m_i+1)^+
(m_{i+1},m_{i+1}+j-2)^+]]\label{L}
\end{equation} 
In (\ref{L}) the complicate positive root of $I^{(+1}_i$ subspace is
presented in form of the consequent commutator of the third positive 
roots from which it is composed. The third order commutator 
in the case, when one ( or two simultaneously) of its component goes
to
unity ($s=R_i+1\to (m_i+R_i,\bar m_i)^+=1; j=1\to
(m_{i+1},m_{i+1}-1)^+=1$) 
is reduced to usual commutator of two roots or disappeared at all.

And so coefficient function $\bar P^i$ is described with the help of 
$(R_i+1)\times (R_{i+1}+1)$ rectangular matrix.

$L^-_i=(L^+_i)^T$ with the simultaneously exchange the coefficient
matrix
$\bar P^i(y)$ on $(R_{i+1}+1)\times (R_i+1)$ rectangular matrix
$P^i(x)$.
We call this operation as "hermitian conjugation".

After these necessary explanations we are able to pass to concrete 
calculation of (\ref{MCC}) for $m_i$ red algebra.

As was mentioned above different from zero input in this expression
may give 
only $L^{\pm}_i,L^{\pm}_{i-1}$ components of operators $L^{\pm}$. So 
the four terms from the possible combinations of two last objects
arise.
As it will be clear from what follows the reciprocity combinations 
$L^{\pm}_i,L^{\mp}_{i-1}$ lead to zero result and so finally in
(\ref{MCC})
it remains the sum of two terms $L^+_i,L^-_i$ and
$L^+_{i-1},L^-_{i-1}$.

All composite roots of the generator $L^-_i$ are constructed from the
negative simple roots (generators) $X^-_j$ with the index $j$ no less
than
$m_i$. The bases vectors of the form:
$$
X^-_j X^-_j....X^-_m\ve{m},\quad X^-_j X^-_{j+1} X^-_j....X^-_m\ve{m}
$$
in the case of $A$ series have the zero norm and so in the result of
the 
action of $L^-_i$ operator on each basis vector of (\ref{SV1}) 
only the vector states of the same kind may arise but with more
number of 
lowering operators with consequent raising on unity indexes. The same
is true
up to operation of "hermitian conjugation" with respect to action of
$L^+_i$
on bra state vectors of the first fundamental representation of $m_i$
red algebra.

The result of the action of $L^+_i$ on $j$ state vector from
$(\ref{SV1})$
may be represented in the form:
$$
\vc{t} L^+_i=
$$
\begin{equation}
\vc{m_i}(m_i,m_i+t-2)^+L^+_i=\vc{m_i}(m_i,\bar m_i+1)^+
\sum_{j=1}^{R_{i+1}+1}  \bar P^i_{t,j}
(m_{i+1},m_{i+1}+j-2)^+)\label{LA}
\end{equation}
The result of action of $L^-_i$ on a ket state vector $\vc{t}$ from
the left
arise as "hermitian conjugation" of (\ref{LA}). 

After this in the process of calculation of (\ref{MCC}) operators in
sum of 
(\ref{LA}) and analogical sum of negative roots generators from the
right may 
be taken out from determinant sign in the form of generators of the
right 
(left) of regular representation shifts ( see (\ref{AM})). The
remaining 
determinant satisfy all conditions of the generalised Jacobi identity 
(\ref{GYI}) with the system of the state vectors (\ref{SV1}) added by
the 
vector $\ve{m_i+R_i}\equiv X^-_{m_i+R_i} \ve{R_i+1}$. After
calculation the 
values of Cartan generators on it we obtain:
\begin{equation}
Det_{R_i+2}=< b_{i-1}) >^{R_i+1} \vc{m_{i+1}} K \ve{m_{i+1}}
\label{VG}
\end{equation}
The generators of right (left) regular representations in each terms
contain
positive (negative) simple roots beginning exactly from index
$m_{i+1}$.
This leads after the action on the highest vector of first
fundamental 
representation of the $m_{i+1}$ red algebra (\ref{VG}) to the matrix
elements 
in this representation, which coincide with matrix elements of the
matrix
$u_{i+1}$. Taking into account the value of calculated $Debt u_i$
(\ref{DU}), 
we obtain finally expression for the first term of (\ref{MCC}), which
we 
present in matrix form including in it $u^{-1}_i$ from the left side
of 
(\ref{MCC}):
\begin{equation}
{<b_{i-1}>\over < b_i > } u^{-1}_i \bar P^i u_{i+1} P^i\equiv 
y^{-1}_i \bar P^i y_{i+1} P^i \label{PS}
\end{equation}
where $\bar P^i(y)$ and $ P^i(x)$ are correspondingly $(R_i+1)\times
(R_{i+1}+
1)$ and $(R_{i+1}+1)\times (R_i+1)$ rectangular matrices of
coefficients of 
$L^{\pm 1}_i$ operators.

The calculation of the term with the pair $L^+_{i-1},L^-_{i-1}$ is a
bit 
more complicated. First of all, let us pay attention ton the fact
that each 
generator of the positive complicate root, belonging to
$I^{(+1}_{i-1}$,
may be uniquely presented as a consequent commutator of three terms 
$$
(s,k)^+=[(s,\bar m_{i-1})^+,[(m_i-1,m_i-1)^+,(m_i,k)^+]]
$$
where $(s,\bar m_{i-1})^+$- generator of the positive root of the
last 
column of $m_{i-1}$ red algebra, $(m_i,k)^+$ the element of the first
line
of $m_i$ red algebra and $(m_i-1,m_i-1)^+ \equiv X^+_{m_i-1}$-
generator of the simple black root on the boundary of these two red
algebras.

Except of these regular elements to $ I^{(+1}_{i-1} $ belong also the 
generators:
$$
(m_i-1,m_i-1)^+,\quad [(s,\bar m_{i-1})^+,(m_i-1,m_i-1)^+]\quad,
[(m_i-1,m_i-1)^+,(m_i,k)^+]
$$
composing respectively the first line and column of $ I^{(+1}_{i-1}
$.

Now let us consider the action of the operator $L^+_{i-1}$ on the bra
bases 
vectors of $m_i$ red algebra (\ref{SV1}). We have consequently:
$$
\vc{k} L^+_{i-1}\equiv 
\vc{m_i} X^+_{m_i}...X^+_{m_i+k-2} \sum_{s,j} \bar P^{1,i-1}_{s,j}
[(s,\bar m_{i-1})^+,[(\bar m_{i-1}+1,\bar m_{i-1}+1)^+,(m_i,j)^+]]=
$$
\begin{equation}
-\vc{m_i} X^+_{m_i}X^+_{m_i-1}(m_i+1,m_i+k-2)^+ \sum_{s,j} \bar
P^{1,i-1}_
{s,j}(m_i,j)^+ (s,\bar m_{i-1})^+.\label{NE}
\end{equation}
The same expression up to "hermitian conjugation" arises as a result
of the action of $L^-_{i-1}$ on the ket state vector (\ref{SV1}). Now
the 
process of calculation of determinant (\ref{MCC}) may be performed in
two steps.
All generators of complicated positive (negative) roots may be taken
out from 
the sign of determinant in the form of corresponding generators of
left 
(right) regular representation shifts. The sequence of bases vectors
of
remaining determinant coincide with bra (ket) basis (\ref{SV1}) added
by
state vector $\vc{m_i} X^+_{m_i} X^+_{m_i-1}$. All conditions of
generalized
Jacobi identity (\ref{GYI}) are satisfied and after calculation of
the summed
values of Cartan generators on such sequence of bases vectors we come
to the
following value for determinant from (\ref{MCC}):
\begin{equation}
Det_{R_i+2}= < m_i+R_i >< m_i-1 >^{R_i-1} < m_i-2 > < m_i+1
>\label{VG1}
\end{equation}
In the last expression $< m_i-2 >$ is the matrix element between the
highest vector states of last fundamental representation of $m_{i-1}$
red
algebra; the generators $(s,\bar m_{i-1})^{\pm}$ of left (right)
shifts of
regular representation acting on this function lead to matrix
elements of 
the matrix $u_{\bar m_{i-1}}$ from (\ref{MTE}); $< m_i+1 >$ is the 
highest vector of the second fundamental representation of $m_i$ red
algebra,
which it will be suitable with the help of first Jacobi identity to
present 
in the form:
$$
< m_i+1 >< m_i-1 >= 
$$
\begin{equation}
\vc{m_i}K\ve{m_i}\vc{m_i}X^+_{m_i}KX^-_{m_i}\ve{m_i}-   
\vc{m_i}KX^-_{m_i}\ve{m_i} \vc{m_i}X^+_{m_i}K\ve{m_i}.\label{NAB}
\end{equation}                                       
In (\ref{NE}) there are two types of generators of the 
complicate roots constructed from raising operators $(m_i+1,k)^+$ and 
$(m_i,j)^+$ beginning correspondingly from $m_{i+1}$ and $m_i$ simple
roots
generators. Each one of them give nonzero input under the action only
on the 
one of two factors in (\ref{NAB}). The positive roots generators
$(m_i+s-1,
\bar m_i)^+$ acting from the left on the highest vectors 
$< m_i+R_i >\equiv < \bar m_i>$ ( and the same negative ones from the 
right) lead to the matrix $(t u_{\bar m_i}t)^T$, which with the help
of
(\ref{DU}) is connected with $y_{i-1}^{-1}$.

Gathering all these results together we obtain for the second
part of ${det_{R_i+2}\over det_{R_i+1}}$ from (\ref{MCC}), connected
with
the input of the the pair $I^{(\pm 1}_{i-1}$ the following expression 
(we include in it the $u_i^{-1}$ from the left):
$$
(\ln < i-1 >)_{xy} I-P^{1,i-1} y_{i-1}^{-1} \bar P^{1,i-1} y_i,\quad 
I=u_i^{-1}u_i
$$
or united $u_i$ with $< b_{i-1} >^{-1}$ we obtain 
finally equation, which functions $y_i$ satisfy:
\begin{equation}
(y_i^{-1} (y_i)_x)_y=y^{-1}_i \bar P^{1,i} y_{i+1} P^{1,i}-P^{1,i-1}
y_{i-1}^
{-1} \bar P^{1,i-1} y_i \label{PSS}
\end{equation}
In the particular case of the equal dimensions of all matrices $y_i$
and the  
choice of all $P,\bar P$ as the unity matrix of corresponding
dimension we
come back to the system of equations of matrix Toda chain.

\section{The case of arbitrary graded subspaces involved}

The elements of $I^{(+k}_i$ k-graded subspace may be enumerated in
the same
way as it was done in the case of $I^{(+1}_i$ one. In their
construction the 
following sequence of generators takes part: $(m_i+s-1,\bar m_i)^+$ -
elements 
of the last column of $m_i$ red algebra, $(m_{i+k},m_{i+k}+j-2)^+$ -
the same
of the first line of $m_{i+k}$ - red algebra and generator 
$(\bar m_i+1,m_{i+k-1}+1)^+$ element of the lower left corner of
$I^{(+k}_i$ 
subspace.

The "regular" element of $I^{(+k}_i$ may be presented in form:
$$
(s,j)^+=[(m_i+s-1,\bar m_i)[(\bar m_i+1,\bar m_{i+k-1}+1)^+,(m_{i+k},
m_{i+k}+j-2)^+]],  
$$
$$
1\leq s\leq R_i+1, \quad1\leq j\leq R_{i+k}+1.
$$
About the meaning of the last expression in the case, when one or two
its 
elements formally equal to 1, see the comments after (\ref{L}) of the
previous 
section.

The components of $L^{+k}_i$ operator, belonging to $+k$ graded
subspace, may 
be written in the form very near to (\ref{L}):
\begin{equation}
L^{+k}_i=\sum_{s=1}^{R_i+1} \sum_{\bar j_k=1}^{R_{i+k}+1}
\bar P^{k,i}_{s,j}(y) [(m_i+s-1,\bar m_i)^+[(\bar m_i+1,\bar
m_{i+k-1}+1)^+,
(m_{i+k},m_{i+k}+\bar j_k-2)^+]]\label{Lk}
\end{equation} 

As always:
$$
L^{-k}_i=(L^{+k}_i)^T,\quad \bar P^{k,i}_{s,j}(y)\to P^{k,i}_{j,s}(x) 
$$
And thus the generators $L^{(\pm k}_i$ are described by $(R_i+1\times 
R_{i+k}+1)$ and $(R_{i+k}+1\times R_i+1)$ rectangular matrices of
their 
coefficients.

In concrete calculations below we restrict ourselves by the choice
$m_1=m_2=2
$. The situation in the case of arbitrary $m_1,m_2$ will become clear
and obvious after this consideration.

As one can see the nonzero input under the action on the bases
state vectors of $m_i$ red algebra may be give only the following
components of $L^{\pm}$ generators:
$$
L^{\pm 1,i}+L^{\pm 2,i}+L^{\pm 2,i-2}+L^{\pm 1,i-1}+L^{\pm 2,i-1}
$$

In the process of calculation of the determinant in (\ref{MCC}) the
following
sequences of $R_i+2$ bases vectors may arise: the sequences of the 
$R_i+1$ bases vectors of the first fundamental representation of
$m_i$-th red 
algebra (\ref{SV1}) added by $\vc{m_i} (m_i,\bar m_i+1)^+,
\vc{m_i} (m_i-1,m_i)^,\vc{m_i} (m_i-1,\bar m_i+1)^+$. The
determinants for 
the two first sequences were calculated in the 
previous section (\ref{VG}),(\ref{VG1}). They belong to different
irreducible 
representations and by this reason reciprocity terms between them are 
cancelled. In calculations below reciprocity terms between the 1-3,
(3-1), 2-3,
(3-2) bases take place and calculation of the corresponding
determinants as 
also in the (3-3) case need additional attention.

Let us consider for instance the case (1-3). So we want to calculate
the 
determinant:
$$
Det_{R_i+2}\vc{1} K \ve{3}
$$
With respect to left shifts it is the highest vector of
($Vh_{b_{i-1}}=R_i-1,
Vh_{m_{i+1}}=1$) irreducible representation of the initial $A$
algebra.
Thus $Det_{R_i+2}$ is some matrix element between the highest vector
from the 
left and some other bases vector of this representation from the
right. All 
such vectors may be obtained with the help of the action some number
of the 
lowering operators on the highest vector (see section 2). Calculating
the 
values of Cartan elements from right we uniquely come to conclusion
that this 
is the single lowering generator $X^-_{m_{i-1}}$ and thus:
\begin{equation}
Det_{R_i+2}\vc{1} K \ve{3}={1\over R_i+1} <m_{i+1}> X^-_{m_i-1}
<b_{i-1}>^
{R_i+1} \label{LD}
\end{equation}
With the help arguments of the same kind we obtain:
\begin{equation}
Det_{R_i+2}\vc{2} K \ve{3}=
<b_{i-1}>^{R_i-1} <\bar m_{i-1}>(m_i+1,\bar m_i+1)^- <m_i+1>
\label{LDD}
\end{equation}
And, finally, in the case of $(3-3)$ we have:
$$
Det_{R_i+2}\vc{3} K \ve{3}= 
$$
$$
<b_{i-1}>^{R_i-1}<m_{i+1}>\vc{m_i-1} X^+_{m_i-1} 
K \ve{m_i-1}\vc{m_i-1} K X^-_{m_i-1}\ve{m_i-1}+
$$
\begin{equation}
<b_{i-1}>^{R_i-1}<m_i-2> 
(m_i+1,\bar m_i+1)^+(m_i+1,\bar m_i+1)^- <m_i+1> <\bar
m_i+1>\label{LLD}
\end{equation}
The proof of this formulae see in Appendix I.

Now we would like to calculate the input of $(1-1)$ pair in
(\ref{MCC}).

The action of $L^+_i$ ( two its first terms) generator on state
vectors of 
$m_i$ red algebra takes the form (compare with the same calculations 
(\ref{LA})):
$$
\vc{\bar t} L^+_i=
$$
\begin{equation}
\vc{m_i}(m_i,\bar m_i+1)^+(\sum_{\bar j_1=1}^{R_{i+1}+1}  
\bar P^{1,i}_{\bar t,\bar j_1} (m_{i+1},m_{i+1}+\bar
j_1-2)^++\sum_{\bar j_2=
1}^{R_{i+2}+1}\bar P^{2,i}_{\bar t,\bar j_2} (m_{i+1},m_{i+2}+\bar
j_2-2)^+)
\label{LAL}
\end{equation}

Now we substitute (\ref{LAL}) and "hermitian conjugated" result of
action
$L^-_i$ on a ket state vectors (\ref{SV1}) in (\ref{MCC}). In what
follows
we will not specially remind each time that the same operation is 
necessary to do simultaneously from right also. Taking out of the
determinant 
sign the generators in the circle brackets of (\ref{LAL}) in the form 
of regular representation shifts ( see (\ref{AM})), we come to the
determinant,
which was calculated in the previous section (\ref{VG}). All
generators of the 
composite roots in the circle brackets of (\ref{LAL}) contain
the simple root $X^{\pm}_{m_i+1}$ and thus the determinant
$Det_{R_i+2}$ in
the numerator of (\ref{MCC}) is some linear combination of the matrix
elements 
of the $m_i+1$ - fundamental representation of the initial algebra.

Our nearest goal is to express it in the terms the matrix elements of
$u_i$
matrices and characteristics of the $\pm 1, \pm 2$ graded subspaces.

The equality below, explained and proved in Appendix II, plays the
key role
in what follows (here it is rewritten in the form, necessary for
further 
applications ):
$$
\vc{m_{i+1}} (m_{i+1},\bar m_{i+1}+1)^+K(m_{i+1},\bar m_{i+1}+1)^- 
\ve{m_{i+1}}=
$$
\begin{equation}
(\bar \alpha^{R_{i+1}+1} u_{i+1} \alpha^{R_{i+1}+1})+{\vc{m_{i+1}-1}K
\ve{m_{i+1}-1}\vc{m_{i+2}}K\ve{m_{i+2}}\over
\vc{m_{i+2}-1}K\ve{m_{i+2}-1}} 
\label{BE}
\end{equation}
where $(\bar \alpha^{R_{i+1}+1}, \alpha^{R_{i+1}+1})$ are
$(R_{i+1}+1)$
dimensional line and column vectors with components: 
$$
\bar \alpha^{R_{i+1}+1}_s={\vc{\bar m_{i+1}+1} (m_{i+1}-s+1,\bar
m_{i+1}+1)^+ 
K \ve{\bar m_{i+1}+1}\over< \bar m_{i+1}+1>}.
$$
Complicated roots in the circle brackets of (\ref{LAL}) will be
suitable to 
present in the commutator form:
$$
(m_{i+1},m_{i+1}+\bar j_1-2)^+=[(m_{i+1},\bar
m_{i+1}+1)^+,(m_{i+1}+j_1-1,
\bar m_{i+1}+1)^-],
$$
$$
(m_{i+1},m_{i+2}+\bar j_2-2)^+=[(m_{i+1},\bar
m_{i+1}+1)^+,(m_{i+2},m_{i+2}+
\bar j_2-2)^+]
$$
After substituting the last form ( the generators) of complicate
roots in 
(\ref{LAL}) we pass to the problem of calculations the actions of the 
operators (below and "hermitian conjugated" to it):
\begin{equation}
\sum_{\bar j_1=1}^{R_{i+1}+1} \bar P^{1,i}_{\bar t,\bar j_1}
(m_{i+1}+\bar j_1
-1,\bar m_{i+1}+1)^-+\sum_{\bar j_2=1}^{R_{i+2}+1}
\bar P^{2,i}_{\bar t,\bar j_2} (m_{i+2},m_{i+2}+\bar
j_2-2)^+\label{LAL1}
\end{equation}
on the right hand side of (\ref{BE}).

Thus it is necessary to act by operator (\ref{LAL1}) on both terms of
(\ref{BE}) independently. The second term of (\ref{BE}) is the
product
of matrix elements of highest vectors of different fundamental
representations and so the action on it by each negative root from
left and
positive one from the right is equal to zero. Thus the action of the
first 
sum of (\ref{LAL1}) on the second term of (\ref{BE}) 
leads to the zero output. The action of the second sum is equivalent
to:
$$
{<\bar m_i+1>\over <\bar m_{i+1}+1>}\sum_{\bar j_2,j_2}\bar
P^{2,i}_{\bar t,
\bar j_2} (u_{i+2})_{\bar j_2,j_2}P^{2,i}_{j,t}={<\bar m_i+1>\over
<\bar m_{i
+1}+1>}(\bar P^{2,i} u_{i+2} P^{2,i})_{\bar t,t}
$$
In the process of calculation of the action of operators (\ref{LAL1})
on the 
first term (\ref{BE}) it is necessary to take into account the
following
circumstance. The generators of both sums give different from zero
output,
acting only on the components of the line vector $\bar
\alpha^{R_{i+1}+1}$.
The action of the operators of negative roots from left transform it
to the zero line with single unity on the corresponding place. Action
of the
positive root generators from the second sum of (\ref{LAL1}) leads
to arising of additional rectangular $R_{i+2}+1\times R_{i+1}+1$
matrix $\bar
A^{i+2,i+1}$ with the matrix elements:
\begin{equation}
\bar A^{i+2,i+1}_{\bar j,\bar s}={\vc{\bar m_{i+1}+1} (m_{i+1}+\bar
s-1,
m_{i+2}+\bar j-2)^+K \ve{\bar m_{i+1}+1}\over \vc{\bar m_{i+1}+1} K
\ve{\bar m_{i+1}+1}}\label{AAM}
\end{equation}

Gathering all results above, we obtain the following expression for  
the action the operators (\ref{LAL1}) on the first term of
(\ref{BE}):  
\begin{equation}
(\bar \pi^{1,i} u_{i+1} \pi^{1,i})_{\bar t,t},\label{OM}
\end{equation}
where: 
$$
\bar \pi^{1,i}=\bar P^{1,i}+\bar P^{2,i} \bar A^{i+2,i+1},\quad
\pi^{1,i}=P^{1,i}+A^{i+1,i+2} P^{2,i}.
$$
Taking into account all scalar factors ( the matrix elements of the
highest 
vectors of different fundamental representations), we obtain finally 
expression for output into (\ref{MCC}) the terms, connected with the
pair
graded subspaces $(L^{\pm 1}_i +L^{\pm 2}_i)$:
\begin{equation}
y_i^{-1}(\bar \pi^{1,i} y_{i+1} \pi^{1,i}+\bar P^{2,i} y_{i+2}
P^{2,i})
\label{MID}
\end{equation}

Now we would like to calculate the output of reciprocity (1-3) and
(3-1) 
terms. The general scheme of these calculations remains the same as
above with 
only one difference to use the reciprocity determinant (\ref{LD})
instead of 
(\ref{VG}). Result consists in two additional terms:
$$
-\bar A^{i,i-1}\bar P^{2,i-1}y_{i+1}\pi^{1,i}-\bar \pi^{1,i}y_{i+1} 
P^{2,i-1} A^{i-1,i}
$$
The account of the first part of $(3-3)$ terms, connected with the
first term in 
of determinant formulae (\ref{LLD}) leads to result, which together
with of 
the last terms and (\ref{MID}) may be united in the finally
expression, which 
is coincided with (\ref{MID}) by the form but with:
$$
\bar \pi^{1,i}=\bar P^{1,i}+\bar P^{2,i} \bar A^{i+2,i+1}-\bar
A^{i,i-1}
\bar P^{2,i-1},
$$
\begin{equation}
\pi^{1,i}=P^{1,i}+A^{i+1,i+2} P^{2,i}-P^{2,i-1} A^{i-1,i}.\label{P}
\end{equation}
In what follows under $\pi$ we will always understand the last
definition (\ref{P}).

Now we pass to calculation of the second pair of terms (2-2), giving
output in
(\ref{MCC}). The situation become more clear after presenting in the 
explicit form $\sum_{r=1}^2 L^{\pm r}_{i-r}$:
$$
\sum_{r=1}^2 \sum_{\bar s=1}^{R_{i-r}+1}\sum_{\bar j_r}^{R_i+1}
\bar P^{r,i-r}_{\bar s,\bar j_r}[(m_{i-r}+\bar s-1,\bar m_i)^+[(\bar
m_{i-r}+1
,\bar m_{i-1}+1)^+,(m_i,m_i+\bar j_r-2)^+]] 
$$
After action by the last sum on $\bar t$ - bra state vector of $m_i$
red algebra 
we obtain:
\begin{equation}
\vc{m_i} (m_i-1,m_i)^+(m_i+1,m_i+\bar t-2)^+ \sum_r\sum_{\bar s,\bar
j_r} 
\bar P^{r,i-r}_{\bar s,\bar j_r}(m_{i-r}+\bar s-1,\bar m_{i-1})^+ 
(m_i,m_i+\bar j_r-2)^+.\label{NNE}
\end{equation}
Further evaluation of the last expression exactly coincides with the 
transformation of (\ref{NE}). The difference consists only in
arising 
the additional sum on graded index $r$ instead of only one term $r=1$
in 
(\ref{NE}). To this sum it is applicable recurrent procedure of the
present 
section and finally (\ref{NNE}) may be rediscovered as the following:
$$
\tilde {(\ln < b_{i-1} >)}_{x,y}I-\sum_r \Pi^{r,i-r} y_{i-r}^{-1}\bar
\Pi^{i,i-r} y_i
$$
where now:
$$
\bar \Pi^{1,i}=\bar P^{1,i}-\bar A^{i,i-1}\bar P^{2,i-1} 
\quad \Pi^{1,i}=P^{1,i}- P^{2,i-1}A^{i-1,i}.
$$
and by symbol $\tilde {}$ we have denoted that part of the second
derivative 
of $\ln$ function, which is connected with the output of the
corresponding 
grading subspaces.

Taking into account the reciprocity terms (2-3) ,(3-2) and the part
of the 
terms connected with second term in (\ref{LLD}), we reconstruct $\Pi$
up 
to the $\pi$ from (\ref{P}) and come to the final form
of equation of equivalence for matrices functions:
\begin{equation}
(y_i^{-1} (y_i)_x)_y=y^{-1}_i \sum_r^2\bar \pi^{r,i} y_{i+r}
\pi^{r,i}-
\sum_{r=1}^2 \pi^{r,i-r} y_{i-r}^{-1} \bar \pi^{r,i-r} y_i
\label{PPSS}
\end{equation}
where $\pi^{r,j},\bar \pi^{r,j}$ are defined in (\ref{P}). 

To close the system of equalities it is necessary to calculate the
derivatives of $\pi^{r,j},\bar \pi^{r,j}$ matrix functions with
respect to $y,x$ arguments respectively.

For this purpose it is necessary to calculate the derivatives of
$\bar A_x$ and $A_y$. From the results of section 3 and the
definition of $A,\bar A$ matrix functions it follows that
$$
(\bar A^{i+1,i}_{j,s})_x=(m_{i-1}+s-1,m_{i+1}+j-2)^+L^- \ln <\bar
m_i+1>.
$$
Only three graded components of $L^-$, namely $(L^{-1}_i$,
$L^{-2}_i$, $L^{-2}_{i-1})$, give input in this derivative.
Calculation using only the technique of the present section leads to
the result:
\begin{equation}
\bar A^{i+1,i}_x=y_{i+1}\pi^{1,i}y^{-1}_i,\quad A^{i,i+1}_y=y^{-1}_i
\bar \pi^{1,i}y_{i+1}\label{AF}
\end{equation}

With the help of the last expressions we close the system of
equations of equivalence:
$$
\bar \pi^{1,i}_x=\bar P^{2,i}
y_{i+2}\pi^{1,i+1}y^{-1}_{i+1}-y_i\pi^{1,i-1}
y^{-1}_{i-1}\bar P^{2,i-1},\quad \bar \pi^{2,i}\equiv \bar P^{2,i},
$$
\begin{equation}
\pi^{1,i}_y=y^{-1}_{i+1}\bar \pi^{1,i+1} y_{i+2}P^{2,i}-P^{2,i-1}
y^{-1}_{i-1}
\bar \pi^{1,i-1} y_i,\quad \pi^{2,i}\equiv P^{2,i}\label{FAF}
\end{equation}

The generalization of above calculations to the case of the presence
of graded subspaces of arbitrary dimension is not a very cumbersome
problem. Firstly let us mark that output of the terms of the $k$-th
graded subspace under the action on the bases vectors of $m_i$-th red
algebra consists of $k+1$ terms of $L^+$, which may be presented in
the form:
\begin{equation}
L^{+k}_i=\sum_{p=0}^k \sum_{\bar s_p}^{R_{i-p}+1}\sum_{\bar j_p=1}^
{R_{i+k-p}+1}\bar P^{k,i-p}_{\bar s_p,\bar j_p} (m_{i-p}+\bar
s_p-1,m_{i+k-p}+
\bar j_p-2)^+\label{Lkk}
\end{equation} 
{}From the last expression reader can see that except of the terms
$p=0,p=k$
the generator of each complicate positive root (under the action on
the bases 
of  $m_i$-red algebra) contain as a composite
part the generator $(m_i-1,\bar m_i+1)^+$, which can't be taken out
of the 
sign of determinant in (\ref{MCC}). The terms with $p=0$ as a
component  
contain generator $(m_i,m_i+1)^+$. The terms with $p=k$ contain as
the same 
component $(m_i-1,m_i)^+$. Thus in the general case in the process of 
calculation of determinant (\ref{MCC}) may arise only 5 possibilities 
considered above (\ref{VG}),(\ref{VG1}),(\ref{LD}),(\ref{LDD}) and 
(\ref{LLD}).

We will not repeat in details the corresponding calculation for
obtaining the 
equations of equivalence for matrix $y_i$ functions. These
calculations are 
not very cumbersome by their origin but need many place for
consequent
description. We present here only the final result: 
\begin{equation}
(y_i^{-1} (y_i)_x)_y=y^{-1}_i \sum_r^M\bar \pi^{r,i} y_{i+r}
\pi^{r,i}-
\sum_{r=1}^M \pi^{r,i-r} y_{i-r}^{-1} \bar \pi^{r,i-r} y_i
\label{BBEE}
\end{equation}
where now:
\begin{equation}
\bar \pi^{k,i}=\bar P^{k,i}+\sum_{s=k+1}^M \sum_{t=0}^{s-k}(-1)^t
\bar A^
{i,i-t} \bar P^{s,i-t}\bar B^{i+s-t,i+1},\quad \bar A^{i,i}\equiv
1,\quad
\bar B^{i,i}\equiv 1\label{PFS}
\end{equation}
where now rectangular $R_i+1\times R_j+1$, $j\leq i$ matrices $\bar
A,\bar B$
with the following matrix elements:
$$
\bar A^{i,j}_{\bar p,\bar q}={\vc{\bar m_{i-1}+1} (m_j+\bar
p-1,m_i+\bar q-2)
^+ K \ve{\bar m_{i-1}+1}\over \vc{\bar m_{i-1}+1} K \ve{\bar
m_{i-1}+1}}
$$
$$
\bar B^{i,j}_{\bar p,\bar q}={\vc{\bar m_j+1} (m_i+\bar p-1,m_j+\bar
q-2)
^+ K \ve{\bar m_j+1}\over \vc{\bar m_j+1} K \ve{\bar m_j+1}}
$$

The reader can compare the last and follow below expressions with the
same ones in the case of the main grading \cite{LM1} and come to
conclusion
that difference consists in the correct order of the matrix functions 
involved.

The closed system arise after calculation the derivatives of the
$(\pi,
\bar \pi)$ functions with the respect to arguments $(y,x)$
respectively:

\begin{equation}
\frac{\partial \bar \pi^{r,i}}{\partial x}=\sum_{q=1}^{M-r}
(y_i\pi^{q,i-q}y^{-1}_{i-q}\bar \pi^{q+r,i-q}-\bar
\pi^{q+r,i}y_{i+q+r}\pi^
{q,i+r}y^{-1}_{i+r})\label{FFF}
\end{equation}
\begin{equation}
\frac{\partial \pi^{r,i}}{\partial y}=\sum_{q=1}^{M-r}
(y_{i+r}^{-1}\bar
\pi^{q,i+r}y_{i+q+r}\pi^{q+r,i}-\pi^{q+r,i-q}y^{-1}_{i-q}
\bar \pi^{q,i-q}y_i)\label{FF}
\end{equation}

Thus the system of equations (\ref{BBEE}),(\ref{FFF}) and (\ref{FF})
is
exactly integrable and its general solution is given by the formulae
above.

\section{Outlook}

In some sense in the present paper the initial idea of Sofus Lie is
realized.
Firstly the aim of introduction of the continuos groups was connected
with the
hope to obtain the power apparatus for solving of the differential
equations.  

On the example of semisimple groups of the $A_n$ series (with
arbitrary 
grading in it) we have decoded this idea and described in explicit
form the 
exactly integrable systems, general solution of which is possible to
obtain 
with the help and in the terms of group representation theory. 

This is the main output of the present paper. 

Its results explain also the successful approach to generalization
of the 
theory of integrable systems to the case of the operator-valued
unknown 
functions \cite{GC},\cite{MT2}.
In fact results of sections 4 and 5 show, that considered then
systems are 
exactly integrable under arbitrary dimensions of unknown matrix
functions
involved. Thus the problem of the rigorous mathematical description
of the domain of correct definition of all objects taking part in
these  
systems arises. From the physical point of view, of course, the most 
interesting is the continuous limit to have possibility to include
into the 
game the Heisenberg operators (or -in and -out) of the (quantum)
theory of
interacting fields. One can compare this with the canonical
quantization of 
two-dimensional Toda chain \cite{12}.

{}From results of this paper a new look on the problem of
quasi-determinants 
\cite{GR} arises, because on the way the explicit expression
for these objects are found here in terms of group representation
theory.

And the last comment. We have not presented in this paper the general
form of exactly integrable system as functional of the taken grading,
maximal graded subspaces involved and possible canonical form on
them.
All systems of the present paper and also of (\cite{LM3}) 
can be written in terms of invariant root techniques. Of course,
the question about their generalization on the case of Kac-Moody and
quantum algebras arises. In the first case for the main grading it
is possible to obtain the general solution of periodical Toda chain
in the form of absolutely convergent series \cite{SOM}. Each term of
this expansion is fully defined only by the properties of
representations of Kac-Moody algebras. Thus it is likely to expect
that in the case of arbitrary grading (in Kac-Moody case) we will
have to deal with nonabelian periodical Toda systems and the general
solution of such a problem will be possible represent in the form of
absolutely convergent series in terms of representation theory of
Kac-Moody algebras \cite{KM}. The other alternative point of view on
this problem is yet realized in \cite{GC}. Situation in the case of
the quantum algebras is the more intrigued and the author would not
like to do any guess on this subject except of the mention of yet
solved scalar case \cite{12}.

\noindent{\bf Acknowledgements.}

Author is indebted to the Instituto de Investigaciones en
Matem'aticas
Aplicadas y en Sistemas, UNAM for beautiful conditions for his work.
Author freundly thanks N. Atakishiyev for permanent discussions in
the 
process of working on this paper and big practical help in
preparation
of the text to publication.

This work was done under partial support of Russian Foundation of
Fundamental
Researches (RFFI) GRANT N 99-01-00330.

\section{Appendix I}

The additional to bases  vectors (\ref{SV1}) is the vector 
$\vc{m_i} (m_i-1,\bar m_i+1)^+,(m_i-1,\bar m_i+1)^-\ve{m_i}$. Thus
the summed values of Cartan generators on $Det_{R_i+1}$ are the
following:
$$
Vh_{m_i-1}=R_i-1,\quad Vh_{m_i-2}=1,\quad Vh_{m_i}=1,\quad
Vh_{m_{i+1}}=1
$$
Under the action on determinant with the help of generators of the
simple 
roots $X^{\pm}_{m_i-1},X^{\pm}_{\bar m_i+1}$ from left and right
respectively
we come back to determinants calculated in the main text
(\ref{VG}),(\ref{VG1}
). Thus up to the functions annihilated from the right and left by
generators
of the simple roots the answer is known. The summed values of Cartan
generators
on this term are coincided with the presented above. So up to the
numerical
factor for this term we obtain:
$$
<m_i-1>^{R_i-1}<m_{i+1}><m_i-2><m_i>
$$
The numerical factor may be obtain after comparison both sides for
special
choice of $K=1$ as it was done many times before.

\section{Appendix II}

With the help of consequent differentiation of the second Jacobi
identity 
(\ref{J2}) for the case of $A_n$ series ($k_{i,i\pm 1}=-1$), it is
possible 
to prove the following equality: 
$$
\bar \alpha_{m,m+1,...m+k}=(-1)^{k-1}\bar
\alpha_{m+k,...m+1,m}+\sum_{s=1}
^k (-1)^{s+1}\bar \alpha_{m+s,...m+k}\bar \alpha_{m+s-1,...m}
$$
and the same equality for $\alpha$ functions.

With the help of these equalities it is possible to lead the
following 

$$
\sum_{s=0}^k{<m+s>\over <m+s-1>}\bar
\alpha_{m+s,...m+k}\alpha_{m+k,...
m+s}=< m-1 >^{-1}\bar \alpha^{k+1} u_m \alpha^{k+1}
$$
where $u_m$ is $(k+1)\times (k+1)$ matrix in the basis (\ref{SV1}); 
$\bar \alpha^{k+1}, \alpha^{k+1}$ are $(k+1)$ - dimensional line
(column) 
vectors with the components $\alpha^{k+1}_s=(-1)^{s+1}
\alpha_{m+s-1,...m+k}$; 
the line vector $\bar \alpha^{k+1}$ is "hermitian conjugated" to the
column
$\alpha^{k+1}$ one.

\end{document}